# Gender-based Systematics in HST Proposal Selection

## I. Neill Reid, Space Telescope Science Institute

**Abstract:**

Proposal success rates are calculated for HST Cycles 11 through 21 as a function of the gender of the Principal Investigator (PI). In each cycle, proposals with male PIs have a higher success rate, with the disparity greatest for Cycles 12 and 18. The offsets are small enough that they might be ascribed to chance for any single cycle, but the consistent pattern suggests the presence of a systematic effect. Closer inspection of results from Cycles 19, 20 and 21 shows that the systematic difference does not appear to depend on the geographic origin of the proposal nor does it depend on the gender distribution on the review panels. Segregating proposals by the seniority of the PI, the success rates by gender for more recent graduates (Ph.d. since 2000) are more closely comparable. There is also a correlation between success by gender and the average seniority of the review panel for Cycles 19 and 20, but not Cycle 21. We discuss these results and some consequent changes to the proposal format and additions to the HST TAC orientation process.

1. Introduction

Most observatories allocate telescope time through a competitive peer review process. In the case of Hubble Space Telescope, proposal pressures lead to typical over-subscription factors between 4:1 and 5:1. As in most such processes, members of the HST Telescope Allocation Committee (TAC) are instructed to focus on the scientific merit of the proposals submitted for consideration by each panel. They are asked to set aside any concerns regarding technical issues, such as scheduling or feasibility, unless informed by STScI staff that serious problems exist for a given proposal. Many members of the community implicitly consider proposal assessment to be an objective process, with the best proposals always floating to the top. There is, however, evidence that this is not the case: comparing (independently constituted) preliminary grades from individual HST panel members in past cycles consistently shows significant dispersion, even for the proposals that emerge as the highest ranked for approval. The few instances when cross-comparisons are possible show that different panels can arrive at significantly different rankings for the same proposal set, although the final set of approved proposals is in agreement at the 50-60% level (Hodgkinson, 1997; Brinks et al, 2012).

Time allocation panels generally reach a consensus ranking of the proposals before them, but it is important to recognise that this is a *subjective* process. Each proposal is not just stating a set of scientific objectives, but also making the case why panel members, including some whose expertise lies in different areas, should consider those goals are important, and why a particular proposal team should be given the opportunity to tackle a particular scientific question. In assessing the merit of each proposal, panel members consider not only the science proposed but how the proposal is written, the track record of the proposers, and whether the team has sufficient expertise to carry the proposed research program. Finally, panelists can be influenced by the views expressed by other panel members during the discussion of each proposal.

A number of studies over the last decade or more indicate that unconscious bias (also termed implicit bias) plays a significant role in how committees and individuals deal with a wide range of issues including assessing job applications (Trix & Penka, 2002), evaluating job performance (DiTomaso et al, 2007) and ranking scientific capabilities (Wenneras & Wold, 1997). There are suggestions of systematic differences in writing style that might affect how proposals are received by committees (Argamon et al, 2003). Unconscious bias is generally held to be tied to societal expectations and/or presupposed stereotypes ("schemas") that affect how we judge others (Stewart, 2010). Those biases can be mitigated if they are recognized as likely to be present even in what may seem to be objective processes.

Accepting that proposal assessment is a subjective process does not necessarily lead to the conclusion that unconscious bias plays a role in the final decisions. We can, however, compile statistical information that may provide some insight on the matter. This article examines the success rate of HST proposals as a function of the gender[1] of the Principal Investigator using data from Cycles 11 to 20. For the two most recent cycles, Cycles 19 and 20, we look at the success rate by gender for different geographic regions and for different scientific categories.

[1] Footnote: Gender identity is a complex issue. For the purposes of this article, we consider a binary approach based on external appearance.

2. The HST proposal review

The Hubble Space Telescope is jointly supported by the National Aeronautics and Space Agency (NASA) and the European Space Agency (ESA), and has been operating as a general user observatory since its launch in 1990. Telescope time is available to all members of the international astronomical community, and is allocated in cycles. The typical duration of a cycle is one year, but instrument failures and the timing of servicing missions have extended some cycles. As result, 24 years after launch, Hubble is currently in its 21st observing cycle.

NASA has designated the Director of the Space Telescope Science Institute as the formal allocating official for HST, but in most cases the Director relies on peer-review recommendations made by the Telescope Allocation Committee (TAC). Most proposals are reviewed by topical panels that cover relatively broad scientific areas (e.g. active galactic nuclei and quasars, hot stars and the interstellar medium, exoplanets and solar system science). Each panel has from nine to eleven members, with the number of panels ranging from 10 in Cycle 11 to 14 in Cycle 21. Large programs are reviewed by a super-committee, comprising the panel chairs, 2-3 at-large members and the TAC chair (see Reid, 2013, for more extensive discussion of the evolution of the HST TAC process).

TAC members are drawn from the astronomical community, selected to provide a broad range of expertise capable of assessing the wide variety of programs submitted for consideration. The majority of participants are from US institutions, with ~20% drawn from countries affiliated with ESA and a handful from other countries. Each TAC is constituted anew, although to allow for some institutional memory, typically 25-30% of the reviewers have participated in the previous TAC. The current process involves around 140 participants.

Astronomers can apply for observing time on HST (Guest Observer proposals for orbits, Snapshot proposals for targets) or for funding for Archival Research (AR) or Theory programs. Proposals are assigned to panels based on scientific keywords selected by the proposers, and each panel is assigned a specific number of obits for GO programs. Proposals of all types are reviewed and ranked together, and panelists are asked to focus on recommending what they consider the best science.

Since Cycle 9, the review panels have been structured as pairs or triplets by scientific topic. Thus, major conflicts of interest can be minimized by directing proposals with a panelist as Principal Investigator or co-Investigators to a separate mirror panel with the appropriate expertise. Inevitably, minor conflicts (same institution, close collaboration) remain, and those are dealt with by having conflicted panelists leave the room for discussion and voting on those proposals. The proposals are assessed in a three-stage process: all panelists independently review all the proposals submitted to their panel (through Cycle 21), and, prior to the TAC meeting, submit preliminary grades for proposals where there is no conflict of interest; the proposals are ranked based on averaging the preliminary grades, and the lowest-ranked proposals identified for triage; the remaining proposals are discussed and re-graded in the face-to-face panel meetings, where panelists also have the option of reviving triaged proposals that they feel to merit further discussion; once the grading is complete, the panels review the final list for topical balance. Panel chairs of mirror panels may also consult to consider the overall balance of the portfolio of recommended proposals.

3. **Gender and proposal success rate**

The detailed contents of HST proposals are considered proprietary, and remain so regardless of whether the proposal is accepted or rejected. However, STScI maintains a database with logistical information, including the names and institutions of the Principal Investigator and any co-Investigators. The information in that database serves as the reference for this study. The present analysis focuses primarily on HST proposals submitted for review for Cycles 11 through 20. We also present data for the Cycle 21 review, but consider that separately for reasons discussed further below.

We do not ask proposers to submit gender information as part of their proposal, and generating gender-based statistics has not been a standard part of the post-TAC analysis. Deriving the present set of statistics required a separate effort, identifying the gender of each PI, proposal by proposal. In most cases it is straightforward to draw the appropriate conclusion from the name, and almost all ambiguous cases were resolved through examining publicly-accessible web-based data. The information collected for the present study is not retained within the broader HST proposal database.

HST TAC reviews have been held on an annual basis except when the schedule has been impacted by past servicing missions. Thus, the Cycle 11-20 reviews span the period from November 2001 to May 2012; the Cycle 21 review was held in May 2013. All told, over the 10 cycles from 11 to 20 more than 9,400 proposals were received from the community and ~2100, or ~1 in 4.5, were recommended for acceptance. Over the same period, a total of 792 community members participated in the process, including 600 North American astronomers, 165 ESA astronomers and 27 from other countries. As a comparison, the successful HST Cycle 21 proposals included ~4,000 unique investigators drawn from the worldwide community, notall of whom would be eligible for TAC service; within the US, there are approximately 2,500 faculty members in Astronomy or Astronomy and Physics departments. Thus, the HST TAC samples a minority, but significant minority, of the US astronomical community. It is likely that trends evident in this dataset are broadly representative of at least the US and European astronomical communities.

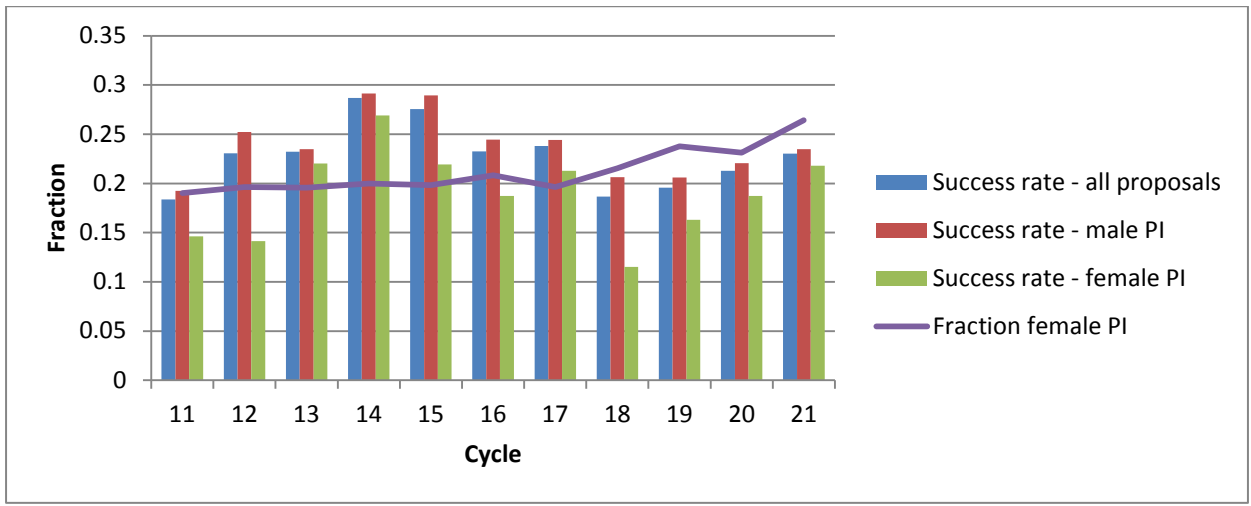

Figure 1: Statistics on the success rate of HST proposals for Cycles 11 through 21. The histograms show the success rates for all proposals, proposals with a male PI and proposals with a female PI; in each case, the statistics encompass all types of proposal (GO, SNAP, AR). The line shows the fraction of submitted proposals with female PIs in each cycle.

### 3.1 Overall statistics

Table 1 summarises the data used in this analysis, listing the total number of proposals submitted and accepted each cycle. The TAC results are shown graphically in Figure 1. For each cycle, we divided the proposals based on the gender of the PI and determined the number of accepted proposals in eachcategory. Figure 1 shows the acceptance rate in each cycle for all proposals and, separately, for proposals with male and female PIs. We also show the fraction of proposals with female PIs for each cycle.

Table 1: Submitted and accepted proposal statistics - Cycles 11 to 20

| Cycle | All | Accepted | Male PI | Accepted | Female PI | Accepted |
|---|---|---|---|---|---|---|
| 11 | 1078 | 198 | 873 | 168 | 205 | 30 |
| 12 | 1045 | 231 | 840 | 212 | 205 | 29 |
| 13 | 905 | 210 | 728 | 171 | 177 | 39 |
| 14 | 725 | 208 | 580 | 168 | 145 | 39 |
| 15 | 737 | 203 | 591 | 171 | 146 | 32 |
| 16 | 821 | 191 | 650 | 159 | 171 | 32 |
| 17 | 958 | 228 | 770 | 188 | 188 | 40 |
| 18 | 1050 | 196 | 824 | 170 | 226 | 26 |
| 19 | 1007 | 199 | 768 | 160 | 239 | 39 |
| 20 | 1085 | 231 | 832 | 183 | 253 | 48 |
| 11-20 | 9410 | 2103 | 7457 | 1750 | 1953 | 353 |
| 21 | 1094 | 252 | 805 | 189 | 289 | 63 |

Overall, the data show two large-scale trends:

- First, the relative number of submitted proposals with female PIs has generally increased over the last ten cycles, rising from ~19% in Cycle 11(~1 in 5 proposals) to 23-24% (almost 1 in 4) in recent cycles. Large programs, reviewed by the TAC super-committee, remain heavily dominated by male PIs, with female PIs typically contributing 1 out of 6 proposals. .
- Second, the success rate for proposals with female PIs is consistently lower than the success rate for proposals with male PIs. The largest discrepancies are in Cycles 12 and 18, when proposals with female PIs had a lower success rate by 11% and 9%, respectively, than proposals with male PIs; the smallest offsets are 1.5% and 2.2% in Cycles 13 and 14, respectively.

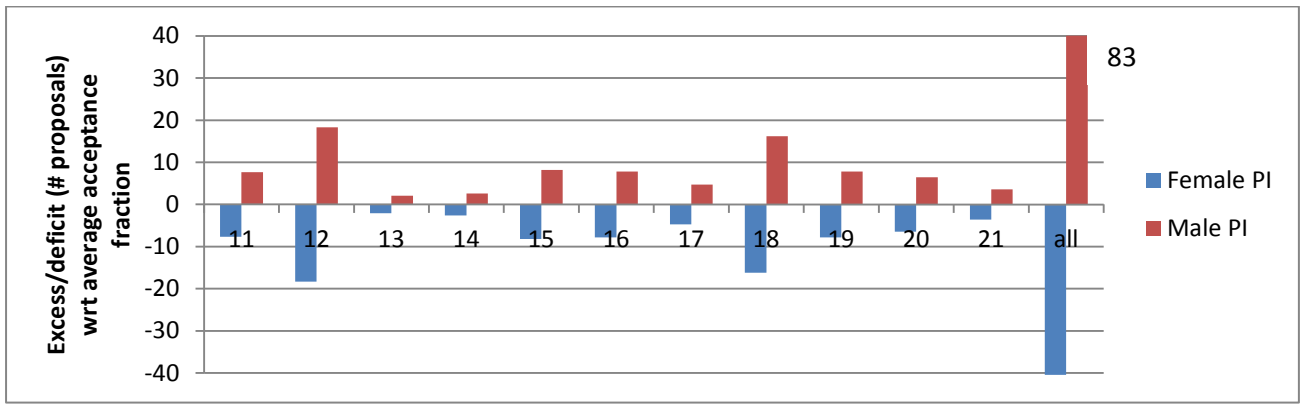

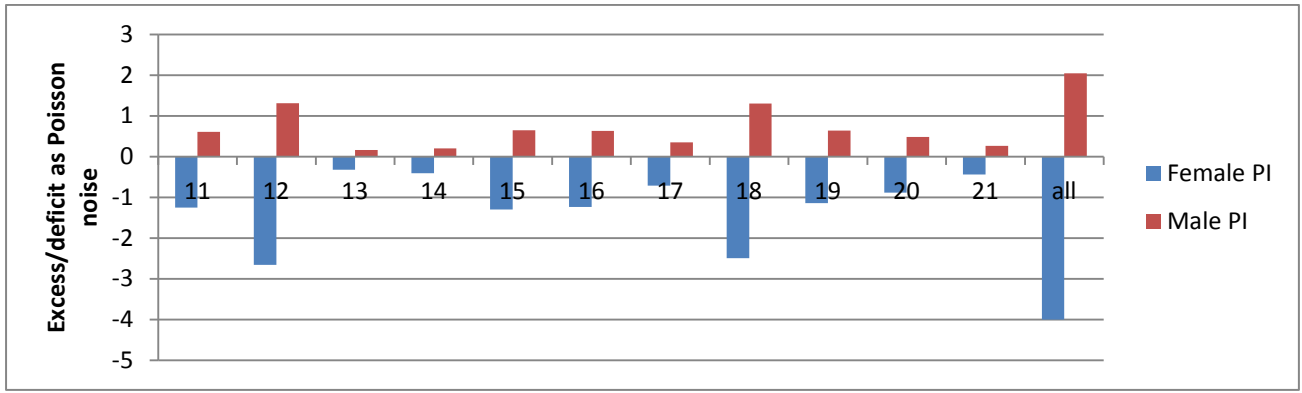

Figure 2: The relative success rates of HST proposals with male and female PIs: the figures show the difference between the actual number of successful proposals each cycle and the expected number based on the overall acceptance rate. The upper figure shows the excess/deficit in proposals for each cycle; the lower panel shows the excess/deficit in Poissonian fashion, i.e. normalized by dividing by the square root of the expected number of accepted proposal. The final columns in each case show the summed statistics for Cycles 11 to 20; note that the y-axis is truncated in the upper plot. The excess/deficit for an individual cycle might be regarded as statistically within the noise, but there is a clear systematic overall pattern.

Figure 2 highlights the systematic pattern present in these data. We use the overall acceptance rate for each cycle to predict the expected number of accepted proposals with male and female PIs. Figure 2 plots the difference between the expected numbers of accepted proposals, $N_e$, and the actual number, $N_a$. We depict the resultant excess or deficit in two ways: as the offset in the absolute number of proposals relative to $N_e$; and as a quasi-Poissonian quantity, normalized by dividing by the square root of $N_e$, the expected number. We show the latter quantity **not** because Poissonian statistics apply to the peer-review process, but because scientists often adopt that perspective in assessing the difference between observations and predictions. Indeed, if one were to consider these results on a cycle by cycle basis, one might be tempted to explain each case as reflecting bad luck or sampling statistics. That argument is more difficult to sustain when viewed in the context of results from all 10 cycles.

In summary, the average success rate for all HST proposals submitted in Cycles 11 to 20 is 22.3%; for proposals with a male PI, the success rate is 23.5%; for proposals with a female PI, the success rate drops to 18.1%. Female PIs lead 20.8% of the proposals submitted for consideration by the TAC, but only 16.9% of the proposals approved for execution.

### 3.2 Selection by geographical origin

We have broken down the Cycle 19 and 20 statistics to determine the success rate for different geographical regions. For this purpose, we segregate proposals in three groups: Europe, including Russia and Israel; North America, including the USA and Canada; and the rest of the world. As Table 2 shows, the overwhelming majority of proposals fall in the first two categories, with approximately 78-80% originating from North America, and 15-18% from Europe. We divide each group in three: accepted proposals; proposals that are triaged based on the preliminary grades, and were therefore not discussed by the panels; and proposals that were rejected after discussion. The triage level was ~35-40% in Cycles 19 and 20.

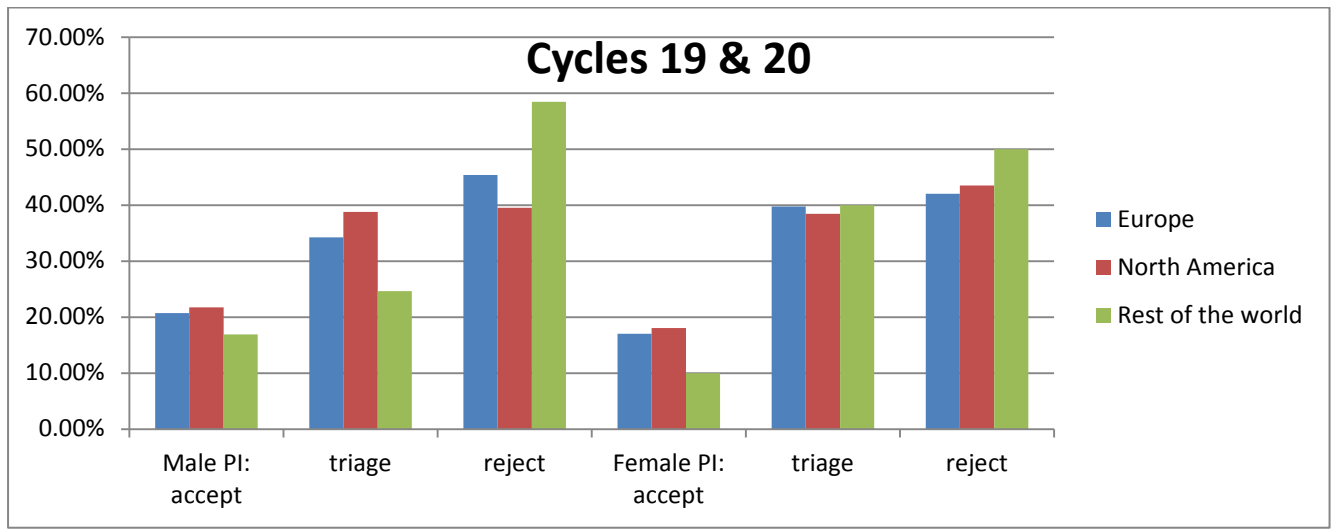

Figure 3: Acceptance/rejection rates for HST proposals in Cycles 19 and 20 as a function of geographical location and PI gender

Table 2: Statistics by geographical region, Cycles 19 and 20

|  | Male PI: all | Accept | Triage | Reject | Female PI: all | Accept | Triage | Reject | F/M ratio |
|---|---|---|---|---|---|---|---|---|---|
| **Cycle 19** | | | | | | | | | |
| Europe | 143 | 27 | 49 | 67 | 45 | 6 | 16 | 23 | 0.31 |
| North America | 595 | 129 | 205 | 261 | 190 | 33 | 67 | 90 | 0.32 |
| Rest of the world | 30 | 3 | 6 | 21 | 4 | 0 | 2 | 2 | 0.13 |
| **Cycle 20** | | | | | | | | | |
| Europe | 162 | 37 | 55 | 70 | 43 | 9 | 19 | 15 | 0.26 |
| North America | 635 | 138 | 272 | 225 | 203 | 38 | 84 | 81 | 0.32 |
| Rest of the world | 35 | 8 | 10 | 17 | 6 | 1 | 2 | 3 | 0.17 |

Figure 3 illustrates the results of this analysis, combining data from both cycles to show the acceptance fraction, triage fraction and rejection (after discussion) fraction by PI gender for each geographical region. The overall European success rate is 20.1%, the North American success rate is 20.9%, and the success rate for the rest of the world is 16%. In all three groups, proposals with male PIs are more successful, with the disparity greatest for proposals from the rest of the world. Segregating by PI gender, proposals from Europe with male PIs had an acceptance rate of 20.9% (64/305), while proposals with female PIs had an acceptance rate of 17.2% (15/88); in absolute numbers relative to the average success rate, this represents an excess/deficit of ~3 proposals. Proposals from North America with male PIs had a success rate of 21.8% (267/1230), while female PIs had a success rate of 18.1% (71/393), an excess/deficit of 10-11 proposals. Finally, proposals from the rest of the world with male PIs had a success rate of 16.9% (11/65), while for female PIs the success rate was only 10% (1/10). Given the small number of proposals in the latter category, the excess/deficit corresponds to less than one proposal, although we note that the offset is in the same sense with female PIs having a lower success rate.

### 3.3 Triage vs. panel discussion

The Cycle 19 and 20 statistics listed in Table 2 provide insight into the gender distribution of proposals that are triaged (i.e. rejected based on independent assessment) and those that are rejected after discussion by the panel. Overall, 37.6% of the proposals were triaged without discussion; those proposals included 38.8% of the proposals with female PIs (190/491) and 37.3% of the proposals with male PIs (597/1599). This corresponds to an offset of 5 proposals; i.e. one would have expected ~184 triaged proposals with female PIs. Focusing only on proposals with North American PIs, the triage fractions are closer, with 38.75% of proposals with male PIs triaged (477/1230) and 38.4% of proposals with female PIs (151/393), where the offset between the two corresponds to an excess/deficit of one proposal, in this case in favour of female PIs.

In comparison, 41.9% of proposals (875/2090) were rejected after panel discussion, including 41.3% of proposals with male PIs (661/1599) and 43.5% of proposals with female PIs (214/491); the latter corresponds to an excess of ~8 proposals relative to the average rejection rate The proportions for proposals originating from North America are 40.5% overall, comprising 39.8% of proposals with male PIs and 43.55% of proposals with female PIs.

### 3.4 Selection by scientific topic

HST proposals are self-classified into scientific categories by the proposer, and this classification determines which panel reviews the proposal. . This provides a means of investigating the relative success of female PIs in a range of research areas. There are six sets of panels, with most panels covering more than category. In those panels, the discussion of a specific proposal is generally led by the experts in that field.

The scientific categories and the associated panels for the most recent cycles are as follows:

- AGN – active galactic nuclei (AGN & quasars)
- QAL – quasar absorption lines (AGN & quasars)
- COS – cosmology (Cosmology)
- CS – cool stars (Stars)
- HS – hot stars (Stars)
- SF – star formation (Planets & star formation/Stars)
- SS – solar system objects (Planets & star formation)

- EXO – exoplanets (Planets & star formation)
- IEG – interstellar medium in external galaxies (Galaxies)
- USP – unresolved stellar populations in distant galaxies (Galaxies)
- ISM – interstellar medium within the Milky Way or nearby galaxies (Stellar populations)
- RSP – resolved stellar populations in nearby galaxies (Stellar Populations)

- Table 3 Gender-based acceptance levels as a function of scientific topic

|  | Male PI: submitted | Male PI: accepted | Female PI: submitted | Female PI: accepted | F/M submitted | F/M accepted | Overall acceptance |
|---|---|---|---|---|---|---|---|
| **Cycles 17-20** | | | | | | | |
| All | 3194 | 701 | 905 | 153 | 0.28 | 0.22 | 20.8% |
| AGN | 358 | 57 | 103 | 17 | 0.28 | 0.30 | 16.1% |
| QAL | 167 | 44 | 52 | 10 | 0.31 | 0.23 | 24.7% |
| COS | 464 | 91 | 111 | 15 | 0.24 | 0.16 | 18.4% |
| CS | 167 | 47 | 50 | 9 | 0.30 | 0.19 | 25.8% |
| HS | 369 | 96 | 74 | 7 | 0.20 | 0.07 | 23.3% |
| SF | 128 | 25 | 72 | 8 | 0.56 | 0.32 | 16.5% |
| SS | 139 | 42 | 37 | 11 | 0.27 | 0.26 | 30.1% |
| EXO | 191 | 43 | 45 | 5 | 0.24 | 0.12 | 20.3% |
| IEG | 118 | 29 | 57 | 13 | 0.48 | 0.45 | 24.0% |
| USP | 421 | 66 | 37 | 11 | 0.38 | 0.45 | 16.8% |
| ISM | 267 | 71 | 58 | 9 | 0.22 | 0.13 | 24.6% |
| RSP | 393 | 88 | 89 | 19 | 0.23 | 0.22 | 22.2% |

We have combined statistics from Cycles 17 to 20 and Table 3 shows the number of submitted and accepted proposals in each category. The final column lists the overall success rate for a given science category; this depends on a number of factors, but most strongly on the resources required for a typical observational program. Solar System and stellar proposals often target bright objects that can yield high quality science in a handful of orbits, while Cosmology proposals generally require observations of very faint objects that demand substantially larger orbit allocations. Thus, the absolute number of proposals accepted, and the corresponding overall success rate, is smaller for Cosmology programs than for Solar System science. The last two columns in the table list the female PI/male PI ratios for submitted and accepted proposals in each category; the difference between these two ratios indicates the relative success rate by gender. Clearly, we should not expect every science category to balance exactly, but it is clear that the larger imbalances are all in the same sense; some science categories (eg EXO, HS, SF) show a success rate for female PIs approximately a factor of 2 below the average value for their male counterparts.

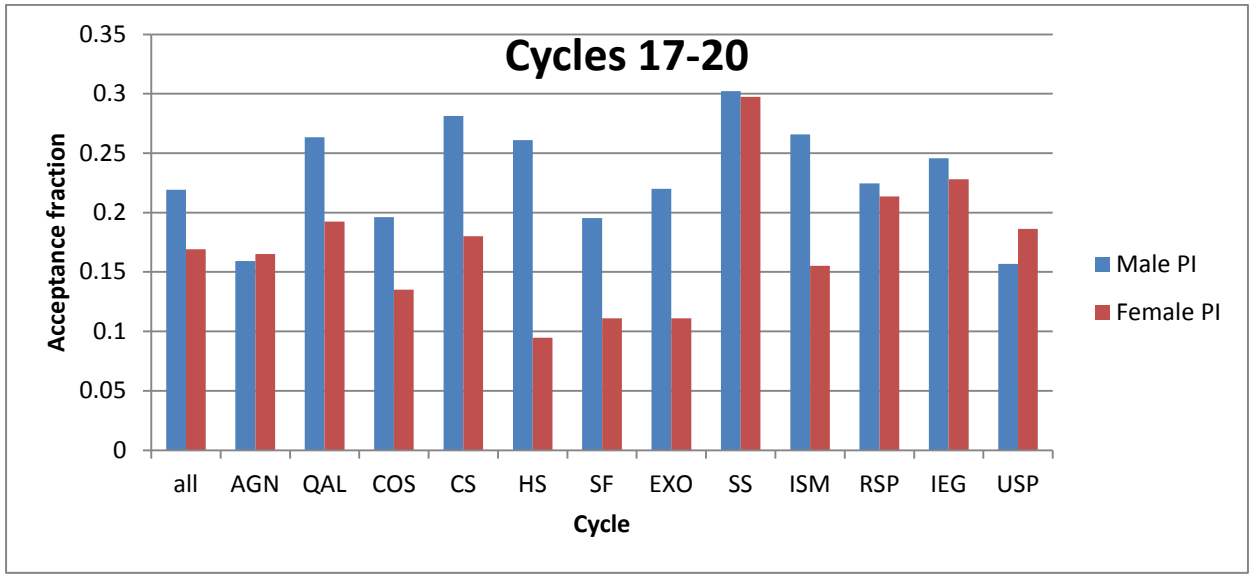

Figure 4: Male/female acceptance rates for HST proposals by scientific category, combining results from Cycles 17 to 20.

Figure 4 plots the acceptance rates by gender within each science category. The overall acceptance rate in these four cycles was 21.9% for proposals with male PIs and 16.9% for proposals with female PIs, but the rates vary substantially among the different science categories. Proposals with female PIs had higher success rates in the AGN and USP categories, while the rates were closer than average for IEG, RSP and SS. However, it is important to recognise that the same panels that produced equitable results for SS proposals showed strong female/male deficits for EXO and SF proposals, while the panels reviewing AGN proposals also reviewed QAL proposals.

Considered by panels, Stars (CS and HS) and Cosmology (COS) show the largest female/male discrepancies, while the Galaxies panels (IEG and USP) have the most equitable results. We should note that there is no significant correlation between the proportion of proposals with female PIs submitted within a science category and the eventual success rate (correlation coefficient $R^2=0.022$); as an example, Star Formation has the highest proportion of submitted proposals with female PIs, but the success rate is close to the Hot Stars science category, which has one of the lowest F/M ratios of submitted proposals in Table 3.

### 3.5. Gender distribution on the selection panels and gender success rates

Does the gender distribution on the Telescope Allocation Committee affect the relative success proposals with male or female PIs? We have made a conscious effort to increase the diversity of the membership of the HST Telescope Allocation Committee over the past ten cycles, and Figure 5 shows that the proportion of female panelists has increased by more than a factor of 2 since Cycle 11. However, increased diversity on the panels has not affected the success rate of proposals with female PIs. We have computed the ratio between the acceptance rates of proposal with female and male PIs for each cycle, and Figure 6 matches that ratio against the female/male gender ratio for the full TAC. Formally, a linear fit to the data shows no evidence for a systematic trend, with a correlation coefficient $R^2=0.03$.

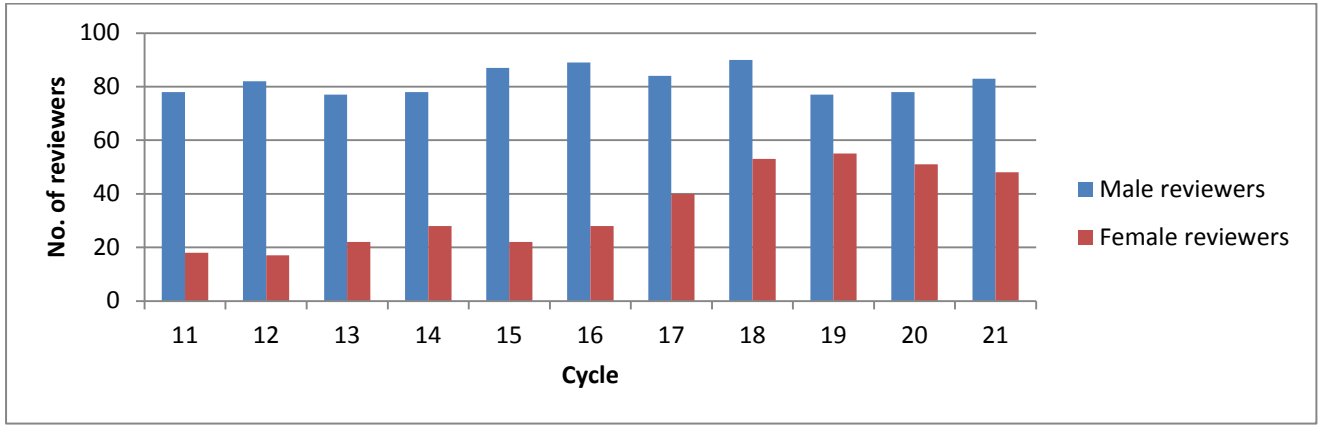

Figure 5: The number of male and female reviewers serving on the HST TAC for Cycles 11 through 21.

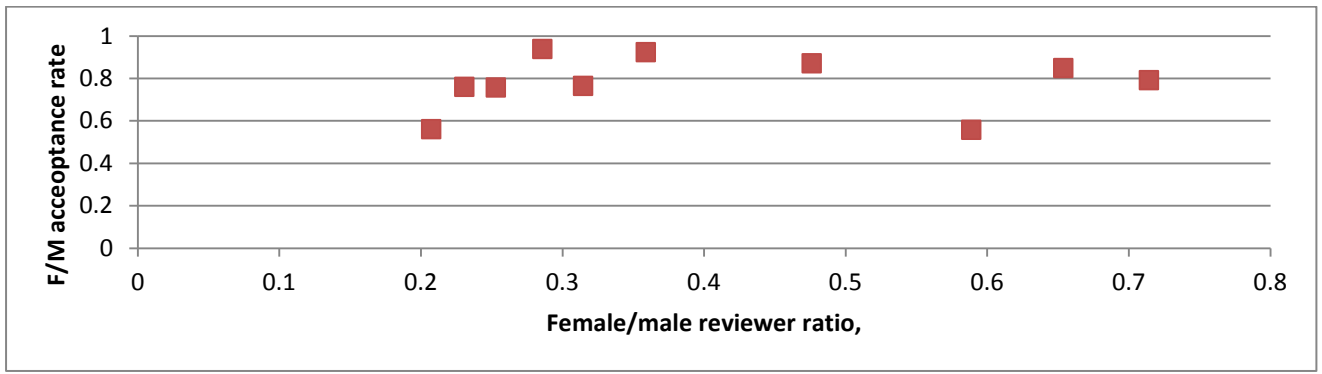

Figure 6: The relative acceptance rate for proposals with female PIs relative to proposals with male PIs plotted against the overall female/male reviewer ratio for Cycles 11 to 20; there is no evidence for a significant correlation.

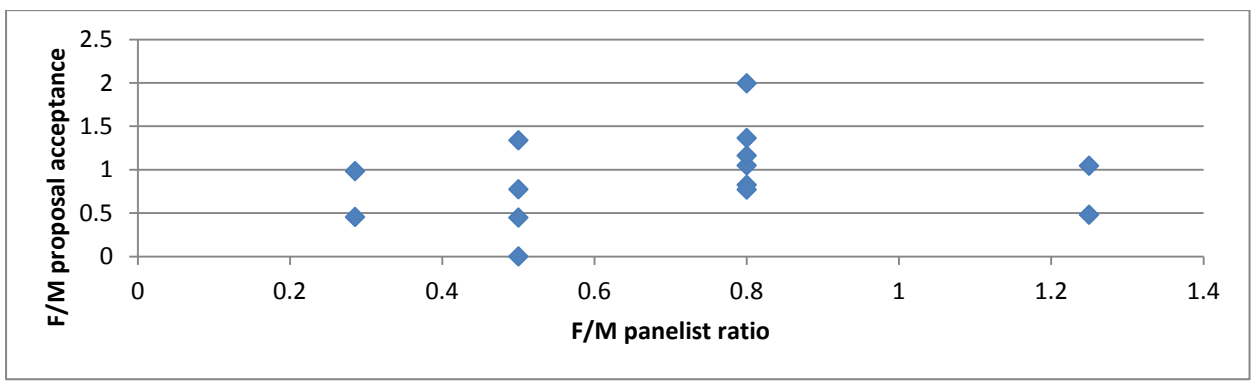

Figure 7: The relative acceptance rate for proposals with female PIs relative to proposals with male PIs plotted against the female/male panelist ratio for the individual panels in Cycle 20; again, there is no evidence for a significant correlation.

Similarly, we have considered the panel-by-panel data from Cycle 20, the cycle with the largest complement of female panelists and the most equitable gender-ratio distribution on the individual panels. Figure 7 matches the relative success rate against the gender composition on the panel. Again, a linear fit to the data shows little evidence for a systematic trend; the formal correlation coefficient, $R^2=0.04$, indicates very low statistical significance.

This result is consistent with previous investigations of unconscious bias. For example, Steinpreis, Anders and Ritzke (1999) circulated CVs for a job applicant and a tenure decision to 238 psychologists; the two sets of CVs were identical except for the gender–specific name assigned to the applicant; male and female psychologists showed a 2:1 preference for the "male" CV in both cases.

### 3.6 Gender demographics

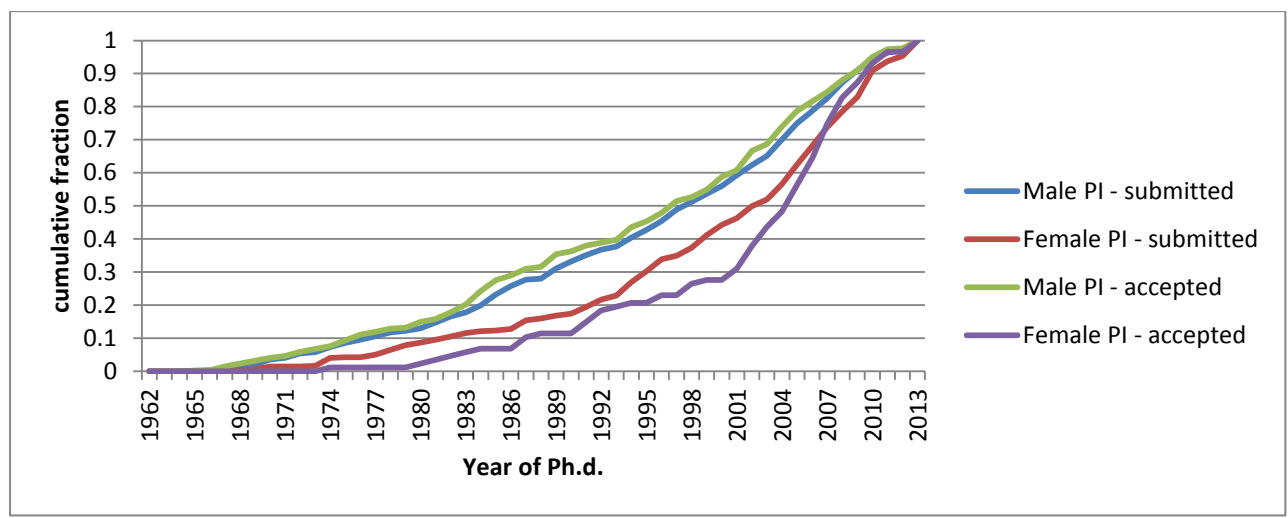

Figure 8: The relative seniority of male and female PIs for Cycle 19 and 20 proposals: we have identified the Ph.d. year (or equivalent) for the PI of each proposal, and show the cumulative distributions, by gender, for submitted and accepted proposals. Note that the submitted and accepted curves for male PIs are closely similar, indicating very similar success rates regardless of seniority; the curves for female PIs are distinctly different, with the larger fraction of accepted proposals from more recent graduates indicating a correspondingly higher success rate.

Experience and reputation are two factors that are often taken into account, either implicitly or explicitly, in assessing research proposals. Indeed, HST proposals include a section where proposers are asked to describe results from previous investigations. We can explore the influence of this component by considering the relative seniority of the Principal Investigators applying for HST time. For each proposal considered by the Cycle 19 or 20 TACs, we have identified the year that a Ph. D. (or equivalent degree) was granted to the Principal Investigator. Figure 8 shows the cumulative distributions for proposals with male and female PIs, plotting data for both submitted and accepted proposals. There are two characteristics to highlight:

> Male PIs of HST proposals include a significantly higher proportion of relatively senior members of the community. The 50-percentile point for submitted proposals lies at 1997/8 for male PIs and at 2003 for female PIs. This undoubtedly reflects the higher proportion of women that have entered the astronomical workforce in recent years.

Perhaps more surprisingly, while the distribution of seniority for accepted proposals is essentially identical to submitted proposal for male PIs, the accepted proposals for female PIs are clearly skewed towards more recent graduates. Seventy-five percent of female PIs on accepted proposals graduated after 1998; that demographic accounts for 60% of the submitted proposals. Male PIs who graduated since 2000 submitted 705 proposals, with 141 accepted for an acceptance fraction of 20%; female PIs from the same "generation" account for 276 submitted proposals and 63 accepted proposals, 22.9%. In contrast, there are 860 proposals from male PIs who graduated between 1962 and 2000; 188 were accepted for an acceptance fraction of 21.9%. Female PIs within the same demographic submitted 203 proposal; only 24 were accepted, an acceptance fraction of 11.8%.

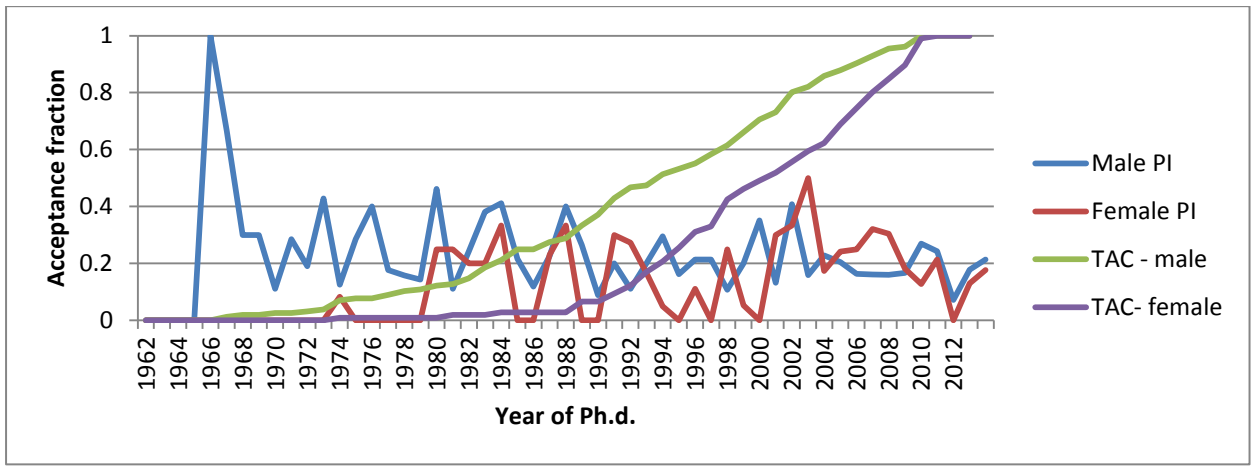

Figure 9: Acceptance fractions as a function of the PI's graduation year for Cycle 19 & 20 proposals. There were no proposals submitted with female PIs prior to 1968 and in 1971, 1972 and 1976; there were no proposals submitted with male PIs in 1963 and 1964; every other year has at least one submitted proposal. The cumulative distributions show the (combined) distribution of seniority for members for the Cycle 19 and 20 TACs, plotting against year of Ph.d. for male (156) and female (106) members.

Figure 9 amplifies on the latter result, plotting the acceptance fraction, year by year, for proposals submitted to HST Cycles 19 and 20. The more equitable acceptance rates for recent male and female graduates are evident. The same figure also shows the seniority distribution for the 262 members of the Cycle 19 and 20 TACs. Both distributions are skewed towards more senior members of the community than the PI distributions, with fifty percent of the male TAC members having graduated before 1993, while the 50-percentile point for female TAC members is 2001.

Figure 9 hints that there may be a tendency for generational differences in the proposal review. To explore that issue further, we have computed the seniority level for each of the Cycle 20 panels by averaging the year of Ph.d. for panel members, without regard to gender. Figure 10 plots that factor against the relative acceptance ratio of proposals with female and male PIs. Clearly, there is no basis whatsoever for assuming a linear relation between these quantities, but a formal least-squares fit does suggest the presence of a correlation. In general, the lower acceptance rates for proposals with female PIs occur on panels with higher average seniority.

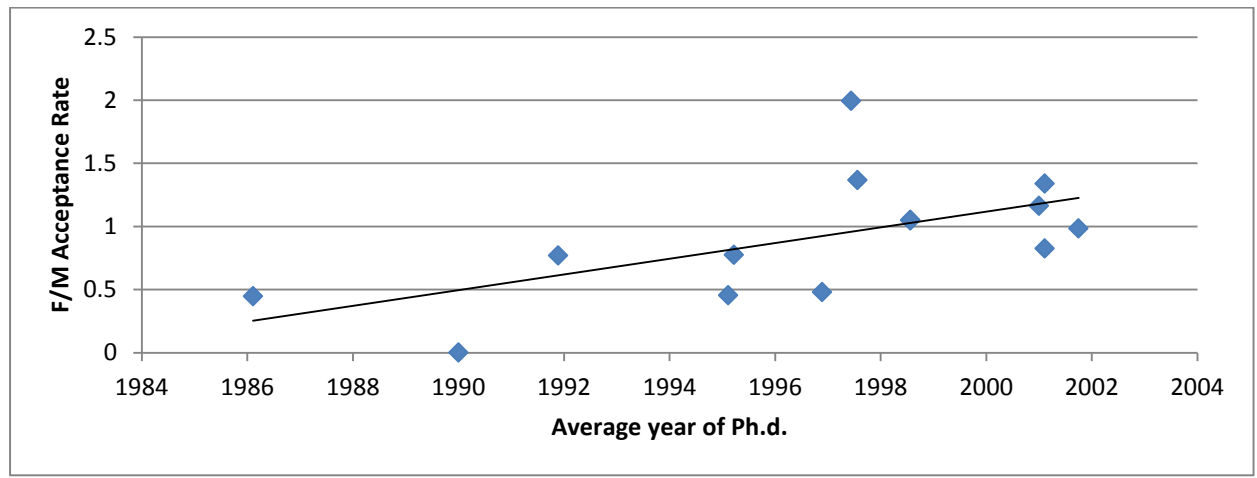

Figure 10: The female/male acceptance ratio for each of the Cycle 20 panels plotted against the average year of Ph.d. of the panelists. A linear fit to the data gives a correlation coefficient of R~0.63.

## 4. The Cycle 21 TAC Results

The results summarized in the previous section were compiled in early 2013, prior to the meeting of the Cycle 21 TAC. Following discussions with the Space Telescope Users' Committee (STUC), it was decided that members of the upcoming TAC should be made aware of the potential for unconscious bias. Consequently, a short discussion was included in the Director's presentation given as part of the general orientation. (The slides are available at http://www.stsci.edu/hst/proposing/panel/peer_review ). Given this procedural change, we consider the results separately from previous cycles.

A total of 1094 proposals were submitted in response to the Cycle 21 Call for Proposals, nine more than in Cycle 20 and the highest number since HST's Cycle 7. These included 805 proposals with male PIs and 289 with female PIs, a ratio of ~2.9:1 as compared with ~3.3:1 in Cycle 20. To date, this is the highest number of proposals with female PIs received in an HST cycle. The proposals were assessed by a Time Allocation Committee that comprised 131 astronomers including 48 female participants (~37%)..

Two hundred and fifty-two proposals were awarded time in Cycle 21, an overall success rate of 23.0%. Those proposals included 189 with male PIs and 63 proposals with female PIs. This cycle therefore has the largest number of successful HST proposals with female PIs, but still represents a lower success rate (21.8%) than for proposals with male PIs (23.5%). In absolute numbers, this corresponds to a deficit/excess of ~4 proposals relative to the average.. As a comparison, in Cycle 20 the success rates were 22.0% for proposals with male PIs and 19.0% for proposals with female PIs (Table 1), corresponding to a ~6 proposal deficit/excess.

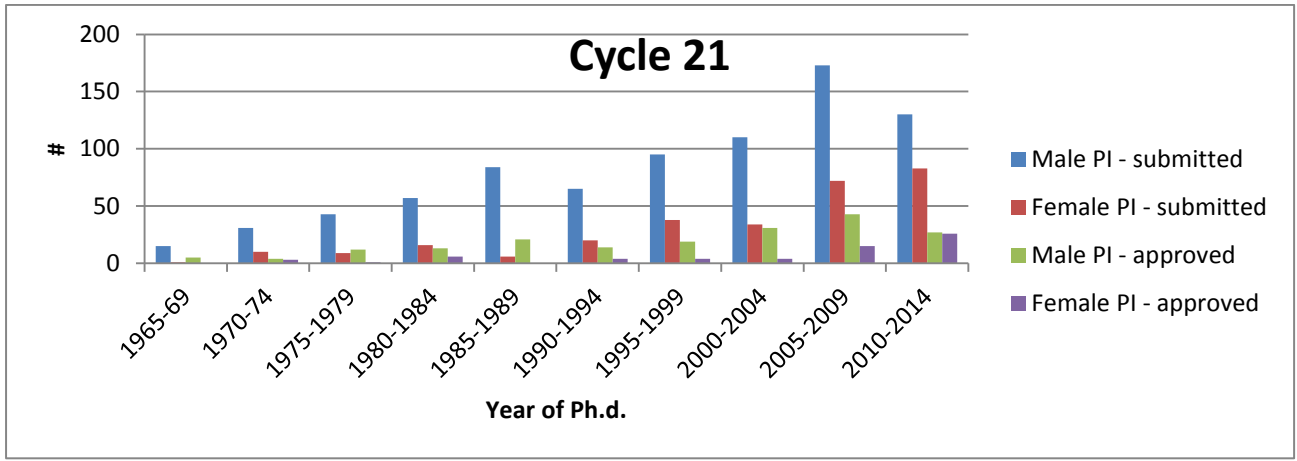

Figure 11: Statistics for submitted and approved proposals from Cycle 21

Figures 11 and 12 show the absolute numbers of proposals submitted and accepted, and the corresponding success rate as a function of the PI's year of graduation. As with Cycles 19 & 20, the success rate by gender is nearer equality for more recent graduates. Specifically, of 805 proposals with male PIs, 400 (52%) are from PIs who graduated (or will graduate) in the 21$^{st}$ century; of the 289 proposals with female PIs, 189 (66%) graduated in the same period. The respective success rates are 100/400 (25%) for male PIs and 43/181 (23.8%) for female PIs, corresponding to a deficit/excess of 1-2 proposals. Among more senior astronomers, 405 proposals were submitted by male PIs who received their Ph.d. between 1962 and 2000; of these, 108 were accepted for a fractional success of 26.7%. The corresponding numbers for female PIs are 108 submitted and 20 accepted proposals for a success rate of 18.5%.

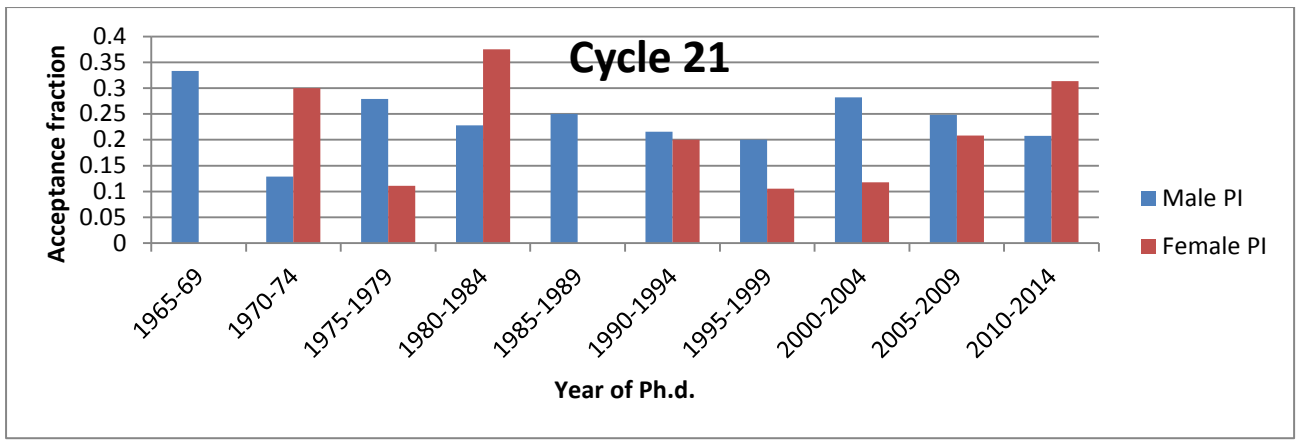

Figure 12: Success rates for male- and female-PI proposals from Cycle 21 matched against the graduation year of the PI

Finally, we have examined the panel data for any evidence of correlations between the relative acceptance rate by gender and either the average seniority or panel gender ratio. Again, the Cycle 21 data show no evidence for a correlation with the latter quantity. In contrast with the Cycle 19 & 20 data, there is also no evidence for a correlation between the relative success rate and the average year of Ph.d. of the Cycle 21 panels (Figure 13).

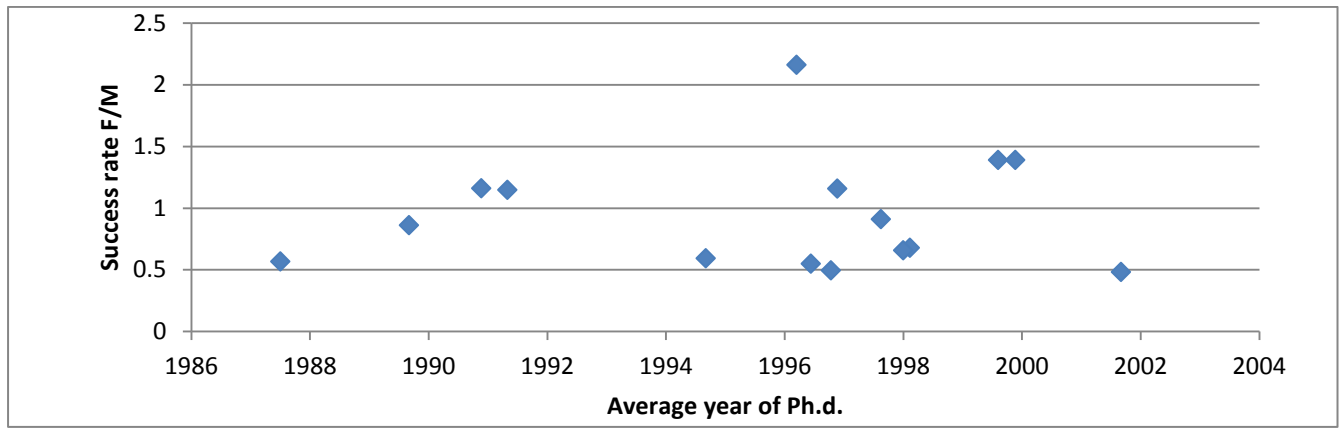

**Figure 13:** The female/male acceptance ratio for each of the Cycle 21 panels plotted against the average year of Ph.d. of the panelists. These data show no evidence for a significant trend.

## 5. Discussion

The statistics presented in sections 3 and 4 indicate that over the past eleven HST cycles a proposal with a male Principal Investigator, on average, stood a greater chance of being approved by the HST TAC than a proposal with a female Principal Investigator. Closer inspection of the results from Cycle 19, 20 and 21 suggests that these results should not be interpreted as a simple across-the-board underlying gender bias. Proposal success rates by gender are more similar for relatively recent graduates, with the corollary that the discrepancy is larger for proposals submitted by more senior female PIs. The data show several other interesting trends: the offset in gender-based results is somewhat smaller for triaged proposals, implying that much of the disparity arises in the discussion phase of the proposal review; changing the gender balance on the panel does not correlate with a changed gender balance of accepted proposals; however, panels with a higher proportion of more junior members tend to produce a more equitable gender distribution; and the same panels can generate different gender-based success rates for proposals within different science categories.

Most tendencies evident in the data are relatively small in scale, implying correspondingly low amplitude in the underlying effect. Nonetheless, the fact that the offsets in the success rates are almost exclusively in the same sense strongly suggests that the effect is real, and that factors other than the assessment of intrinsic scientific value may influence panel decisions. The data at hand do not provide a direct means of identifying those factors. We can only speculate as to the nature of those factors and, in doing so, we should be clear that this is speculation. Moreover, any inferences drawn about possible behavior relate to large-scale group tendencies and should not be taken as characterizing every individual within that group.

Two aspects of the analysis of the Cycle 19 & 20 data merit particular comment: the significantly lower success rate of proposals with more senior female PIs; and the possible correlation between the success rate of proposals with female PIs and the average seniority of the panel. Both effects suggest a generational component contributing to the gender imbalance, either in the proposals themselves or in the proposal review. The former effect is also present in the Cycle 21 data, albeit at a lower level; the latter is not. *Post hoc* is not *propter hoc*, but we note that there was a discussion of the potential for unconscious bias as part of the Cycle 21 orientation[2].

[2]Footnote: After each TAC meeting, a questionnaire is circulated for feedback from the panelists. The Cycle 21 questionnaire included the following question: "At the orientation, Matt Mountain mentioned the issue of unconscious bias and the difference in success rate for male PIs versus female PIs. How did this impact your evaluations?" Ninety-six percent of respondents felt that the discussion had no impact on their assessment of proposals. Of course, the point about unconscious bias is that one does not recognize its presence.

The results plotted in Figure 2 show that the two cycles with the largest male/female disparity in proposal success rates are Cycles 12 and 18. Those two cycles provided the first opportunity to submit HST proposals when the results of a servicing mission were known to the community. Thus, Servicing Mission 3B, which installed the Advanced Camera for Surveys (ACS), was in February 2002, prior to Cycle 11 – but the deadline for Cycle 11 proposals was in September 2001, so the full capabilities of ACS were only known at the September 2002 Cycle 12 deadline. Similarly, while Cycle 17 immediately followed Servicing Mission 4 in May 2009, the Cycle 18 deadline on February 26 2010 was the first time that the community knew that it could take advantage not only of Wide-Field Camera 3 and the Cosmic Origins Spectrograph, but also the refurbished ACS and Space Telescope Imaging Spectrograph. The data in Table 1 show local maxima in the number of proposals submitted in those cycles, indicating higher community interest and consequently greater competition for time. Those circumstances may lead to subtle differences in how panels review proposals: for example, there may be a tendency to favour more established researchers, which, given the demographics, would give and advantage to proposals led by male PIs.

An approach taken in some other fields is to conduct blind proposal review, eliminating all information with regard to the investigators on a given proposal. This approach has the advantage of severely limiting (but not eliminating entirely) the potential for unconscious bias against the proposal team. In the case of HST (and other) observing proposals, this would require a change in approach by reviewers; clearly, if the proposal team is not specified, panelists cannot be asked to consider either their qualifications or their past use of HST (or other facilties) in assessing the current proposal. This would require that proposers adopt a stylistic approach that focuses on the new science being proposed, rather than their past contributions to the subject, so this would require a significant adjustment on the part of the community. One must also take into account the fact that astronomy is a relatively small community, and it may be possible to identify proposers from ancillary data, for example, the list of references.

In any event, the data summarized here highlight a growing change in the overall demographics of the astronomical community. The most recent three HST cycles analysed here have female astronomers leading close to 25% of the submitted proposals, with Cycle 21 attracting the highest number of such proposals to date. One might anticipate (and encourage) a higher proportion of Large and Treasury programs with female PIs in future cycles as those scientists move into more senior positions. The increasing size of that demographic, together with the apparent closer balance in success rates for more recent graduates, may well be the most important factor for change. Achieving an exact balance every cycle may be difficult, but it would be encouraging to have at least one or two cycles where the excess/deficit plotted in Figure 2 becomes a deficit/excess.

## 6. Summary

We have reviewed eleven cycles of the HST Telescope Allocation Process for trends in the proposal acceptance fraction with respect to the gender of the Principal Investigator. The results show that male PIs have a higher success rate than female PIs. The offset might be attributed to sampling statistics for a single cycle, but the consistent results for every cycle indicate that this is a systematic bias. Closer inspection of data from recent cycles shows that the gender disparity is lower among proposals submitted by more recent graduates. There is little evidence for significant variations with regard to geographical origin, and while some science categories give more balanced results than others within the dataset examined, the same panel of reviewers can produce results with significantly different gender ratios for different categories.

Looking forward, we will continue to brief incoming members of the HST TAC on the potential for unconscious bias. Those TAC members are charged with identifying what they consider the "best science" among the proposals submitted to their panel. They will be asked to give careful consideration to the criteria they use in ranking those proposals – the most effective means of countering unconscious bias is recognizing that it may be present. In addition, starting in Cycle 22, we are revising the proposal format: the Principal Investigator is no longer listed on the front page; all investigators are listed together in the proposal, with the PI identified, but giving initials rather than first names. The goal is to provide a broader view of the team who will carry out the research, rather than overly highlighting the PI.

In conclusion, peer review is not a simple objective process; as with any process involving humans, decisions involve a measure of subjectivity. In reviewing a proposal, panelists take into consideration many factors beyond the written word in front of them. It is important that they give due thought to the extent to which external factors might be influencing their final decision, and whether those external factors are actually relevant to the decision at hand.

**Acknowledgements:** Thanks to Sheryl Bruff, Andrew Fox, Andrew Fruchter, Antonella Nota, and Ken Sembach for comments on earlier versions of this paper. Brett Blacker provided the relevant HST proposal information, and Jill Lagerstrom and the STScI library staff tracked down graduation dates for Cycle 19 and 20 PIs.